\documentclass[journal]{IEEEtran}
\usepackage{cite}
\usepackage{hyperref}
\usepackage{makecell}
\usepackage{subcaption}

\usepackage{enumitem} 

\usepackage{amsmath}

\usepackage[version=4]{mhchem}

\ifCLASSINFOpdf
\else
\fi
\usepackage{amsmath}
\usepackage{alphabeta}
\usepackage{graphicx}

\usepackage{siunitx}

\begin{document}
%
\title{LUCAS: A CMOS-Based Fast Readout ASIC for Silicon Photomultipliers – Measurement and Performance Evaluation}
%
%
%

\author{Seyed~Arash~Katourani,~\IEEEmembership{Member,~IEEE}
\thanks{}
\thanks{}
\thanks{}}

%
%

\markboth{}%
{Shell \MakeLowercase{\textit{et al.}}: Bare Demo of IEEEtran.cls for IEEE Journals}
%



\maketitle

\begin{abstract}
This paper presents the design, implementation, and performance evaluation of LUCAS, a low-power, ultra-low jitter ASIC optimized for SiPM readout in Time-of-Flight Computed Tomography (ToF-CT) applications. Leveraging a novel preamplifier design with low input impedance and current-mode operation, LUCAS addresses challenges such as parasitic capacitance of SiPMs and high-speed detection requirements. The ASIC, fabricated in TSMC 65nm CMOS technology, features eight channels with an integrated preamplifier and comparator, achieving an SPTR of 201 ps FWHM at 3.2 mW/ch. Experimental validation includes input impedance measurements, SPTR testing with a pulsed laser source, and energy-to-Time-over-Threshold calibration using monoenergetic X-ray sources. The results demonstrate the effectiveness of LUCAS in enhancing timing resolution while minimizing power consumption, offering significant advancements for high-resolution ToF-CT imaging.
\end{abstract}


%
\IEEEpeerreviewmaketitle

\section{Introduction}

Computed tomography (CT) plays a crucial role in both clinical and pre-clinical medical imaging, offering high spatial resolution, fast scanning, and relatively low costs~\cite{rossignol2024time}. In conventional CT, X-ray photons are emitted from a continuous source, pass through the body, and are detected on the opposite side. Ideally, some of these X-ray photons are absorbed by dense tissues, such as calcified structures, transferring their entire energy. A detector on the opposite side then correlates the level of absorption with tissue density, allowing image reconstruction to produce detailed anatomical images.

However, a phenomenon known as scattering can complicate CT imaging. Scattering occurs when an X-ray photon does not transfer its entire energy to biological structures but instead delivers only part of its energy before deviating~\cite{compton1923quantum}, eventually being detected in a different pixel~\cite{rossignol2020time}. This scattering introduces noise and artifacts into the images~\cite{siewerdsen2001cone}, reducing image quality and making medical diagnoses less certain.

\begin{figure}
    \centering
    \includegraphics[width=0.9\linewidth]{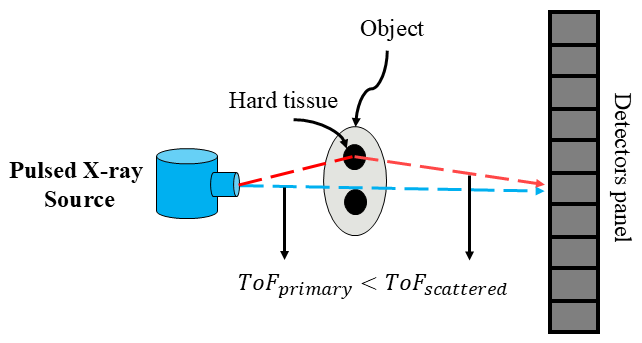}
    \caption{The diagram demonstrates how a pulsed X-ray source, combined with the measurement of X-ray time of flight, can effectively distinguish scatter events from direct primary X-rays. By analyzing the timing information, it becomes possible to separate scattered X-rays from those that travel directly, thereby improving measurement accuracy and reducing noise.}
    \label{fig:PS}
\end{figure}

Methods to address the scattering problem in CT imaging have been extensively discussed in the literature, with a comprehensive review available in~\cite{ruhrnschopf2011general}. These solutions fall into two primary categories:

\begin{enumerate}

    \item Mechanical Approaches: These methods involve physical modifications to the imaging setup aimed at either preventing or reducing the reception of scattered photons. This can be accomplished by adding specific structures to the detector side, such as anti-scatter grids~\cite{bucky1913deactivation} or beam stop arrays~\cite{lazos2008experimental}, or to the source side, like Bowie filters~\cite{boone2010method, graham2007compensators}. Another technique involves adjusting the scanner's diameter using the air gap method~\cite{groedel1926meaning}. However, a significant drawback of these approaches is that they tend to decrease the detector's sensitivity to X-ray photons. Consequently, there's often a need to increase the radiation dose to offset this reduced sensitivity, which is not ideal.

    \item Post-Processing Solutions: This category focuses on estimating the distribution of scattered photons in the final data and then attempting to remove them~\cite{siewerdsen2006simple, ning2002x, jin2010combining, zhu2009scatter}. These approaches are based on estimation, tend to be costly, subject-dependent, and depend on complex computations. 
\end{enumerate}
In 2018, the Université de Sherbrooke introduced a novel approach to address the issue of scattered photons in CT imaging. This method employs high-speed electronics to differentiate scattered photons from primary photons (those that pass through the body without deviation)~\cite{rossignol2020time, fontaine2021pulsed}. Termed "Time of Flight CT" (ToF CT), this technique uses a pulsing X-ray source, rather than a continuous one, and measures the time it takes for the X-ray photons to reach the detector. Scattered photons, due to their deviation caused by the scattering phenomenon, travel a longer distance and consequently have a longer time of flight (Figure~\ref{fig:PS}). By implementing a specific time window, the system can distinguish scattered photons from primary ones, effectively reducing noise and improving image quality~\cite{rossignol2020time}.
As a result, the system's time resolution is directly related to its ability to distinguish between scattered and primary photons~\cite{lemaire2023time}.

Highlighting the significance of time resolution, according to the data published in~\cite{rossignol2020time}, when the system possesses a temporal resolution of 200 picoseconds (ps) at Full Width at Half Maximum (FWHM), there is an observed decrease in the scattered to the primary ratio (SPR) by $60\%$, whereas standard anti-scatter grids (as a commercially available solution) typically achieve a 70 to 95\% 
but at the expense of sensitivity reduction, which decreases to 15 to 30\%~\cite{boone2002development}. As a result, the system's time resolution is directly related to its ability to distinguish between scattered and primary photons~\cite{lemaire2023time}.

From an electronics perspective, the chief impediment to high-resolution detection is the photodetector. Among various mature options, the Silicon Photomultiplier (SiPM) has demonstrated superior timing performance while maintaining acceptable parameters in terms of bias voltage and size. However, the parasitic capacitance of the SiPM complicates fast detection for the readout electronics. This is because the SiPM consists of an array of Single-Photon Avalanche Diodes (SPADs), and the interconnections among them introduce parasitic capacitance, generating a low-frequency pole that limits the system's time resolution.

\begin{figure}[tb]
\centerline{\includegraphics[scale=0.5]{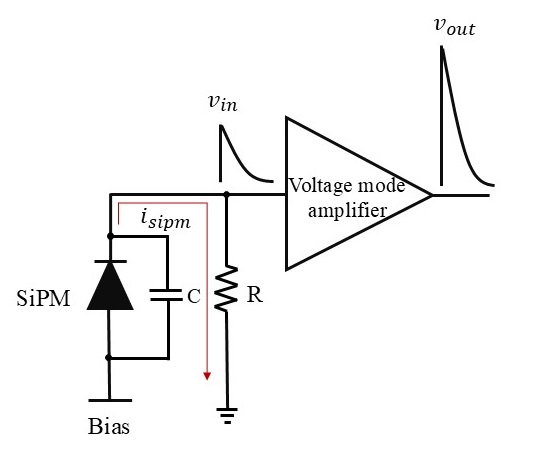}}
\caption{C represents the parasitic capacitance of the SiPM (terminal capacitance), while R denotes the parallel resistor. When a photon strikes the SiPM, it activates the Single-Photon Avalanche Diodes (SPADs) by triggering a breakdown event. This occurs because the SPADs are biased above their breakdown voltage, so the absorption of a photon generates an electron-hole pair, which accelerates and creates an avalanche of charge carriers. This avalanche leads to a large current flowing through the parallel resistor due to its low resistance. The current then generates a voltage across the resistor, which can be amplified by a voltage-mode amplifier. For simplicity, the quenching resistor—essential for stopping the avalanche and resetting the SPAD to its original state—is not shown in the figure.} 
\label{fig:VMA}
\end{figure}

A potential solution involves integrating a low-value resistor in parallel with the SiPM and the readout circuit, thereby shifting the pole to a higher frequency. In essence, most of the current generated by the SiPM is directed through the resistor, and the resulting voltage is amplified by the amplifier (Figure~\ref{fig:VMA}). 

\begin{figure}[tb]
\centerline{\includegraphics[scale=0.5]{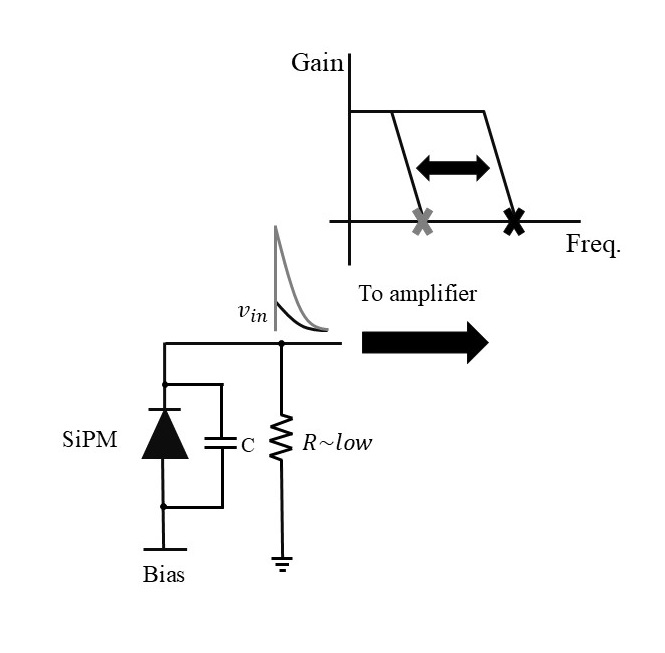}}
\caption{Reducing the value of the parallel resistor shifts the low-frequency pole to higher frequencies, improving the speed of detection by allowing faster signal transitions. However, this reduction in resistance also decreases the input voltage signal, which in turn requires a high-gain amplifier. To achieve this high gain, the amplifier must be carefully designed so as not to introduce any internal poles that could limit the system's bandwidth, as internal poles would reduce the overall frequency response and compromise the high-resolution detection.} 
\label{fig:shift}
\end{figure}

This method, however, presents a trade-off between the necessity for a lower-value resistor for swift detection and the amplifier gain~\cite{calo2019sipm} (Figure~\ref{fig:shift}). To elaborate, reducing the resistor value diminishes the input voltage, necessitating a higher gain for the amplifier, which inversely correlates with bandwidth. Addressing the bandwidth issue requires increased power consumption, imposing limitations on the augmentation of channel count for the scanner level.

\begin{figure}[tb]
\centerline{\includegraphics[scale=0.5]{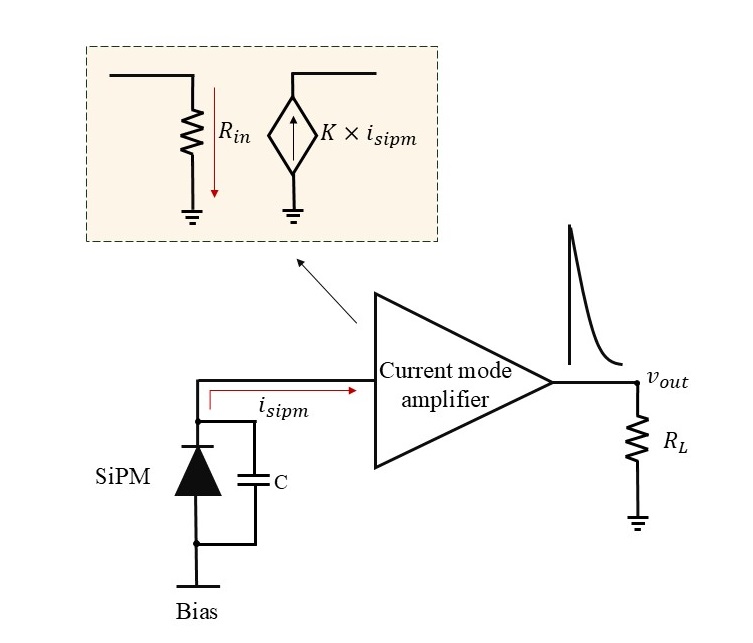}}
\caption{A simplified model of a current-mode amplifier serves as a readout for the SiPM. The SiPM's current is drawn by the amplifier's low input impedance, amplified by a factor $K$, which represents the gain of the current-mode amplifier, and then converted to voltage by the load resistor $R_L$. Additionally, the low input impedance of the amplifier shifts the input pole to higher frequencies.
} 
\label{fig:modelcurrent}
\end{figure}

To address the issue described, a low-input impedance amplifier operating in current mode was developed. This design leverages analog circuit techniques to achieve low input impedance, effectively neutralizing the impact of the Silicon Photomultiplier's parasitic capacitance (Figure~\ref{fig:modelcurrent}). Furthermore, since the amplifier functions in the current mode, it avoids generating a high-impedance node, thereby preventing the creation of a low-frequency pole.

In the following paper:

Section~\ref{sec:ASIC} provides a detailed overview of the chip, discussing its structure and operational principles.

Section~\ref{sec:MM} outlines the materials used in various experiments aimed at characterizing the chip's performance, including its role as a fast readout for SiPMs and as a readout electronic component in ToF-CT projects. The section also details the methodology, explaining how the experiments were conducted and how the results were obtained, particularly in relation to the chip's timing performance, and its functionality in the ToF-CT applications.

Section~\ref{sec:RES} presents the results of each experiment, offering a comprehensive analysis of the chip's performance under various experimental setups and conditions. It highlights the key findings, emphasizing how the chip behaved in different scenarios, including its efficiency, accuracy, and reliability as a fast readout for SiPMs and its application in ToF-CT projects. The section provides a detailed comparison of results across multiple conditions, giving insight into the strengths and potential limitations of the chip's design.

Section~\ref{sec:CON} is dedicated to concluding remarks and discussing future work intended to improve performance.

\section{LUCAS}
\label{sec:ASIC}

\begin{figure}[htbp]
\centerline{\includegraphics[scale=0.2]{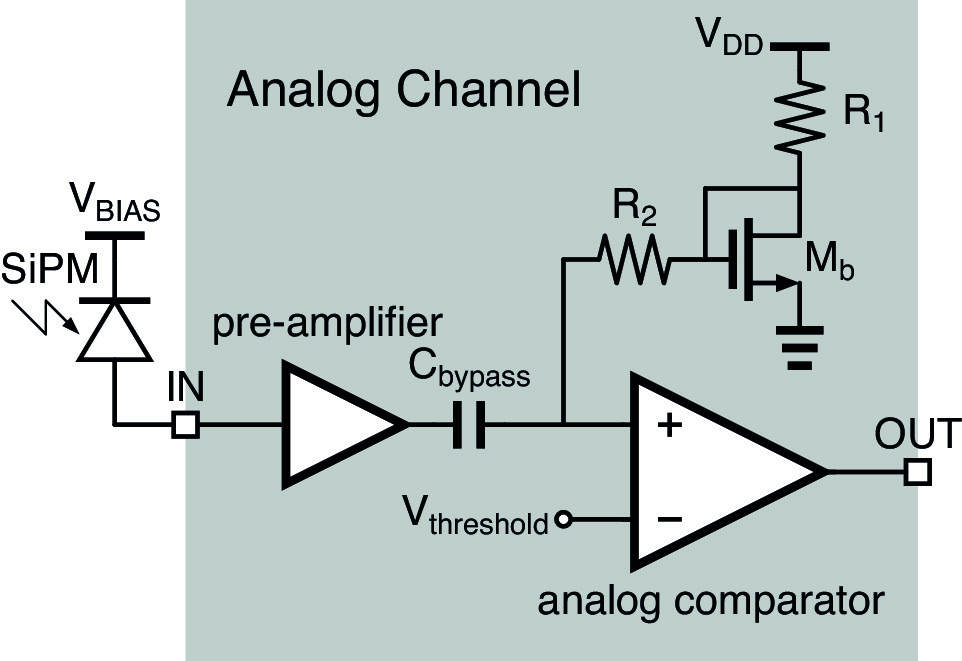}}
\caption{Schematic of a single channel, featuring a pre-amplifier and comparator. The SiPM signal is processed through the low-input impedance pre-amplifier, which facilitates current-to-voltage conversion, followed by a voltage comparator for gain enhancement and digitization. A bypass capacitor, \( C_{bypass} \), ensures uniform input to the comparator across channels. The comparator's input is biased by a large-size diode-connected transistor, \( M_b \), to prevent mismatch effects.} 
\label{fig:CH}
\end{figure}

The Lucas chip, designed with 8 channels, serves as the first electronic prototype and marks the initial steps in building a front-end system for the ToF-CT project. The architecture of each channel, as shown in Figure~\ref{fig:CH}, includes a low-input impedance amplifier (or pre-amplifier), designed to counteract the SiPM capacitance and enable current-to-voltage conversion. The design also incorporates a voltage comparator to enhance gain and support digitization, along with additional circuitry between the pre-amplifier and the comparator to address mismatches and enabling correct biasing, which will be detailed later in this section.

\begin{figure}[htbp]
\centering
\includegraphics[scale=0.4]{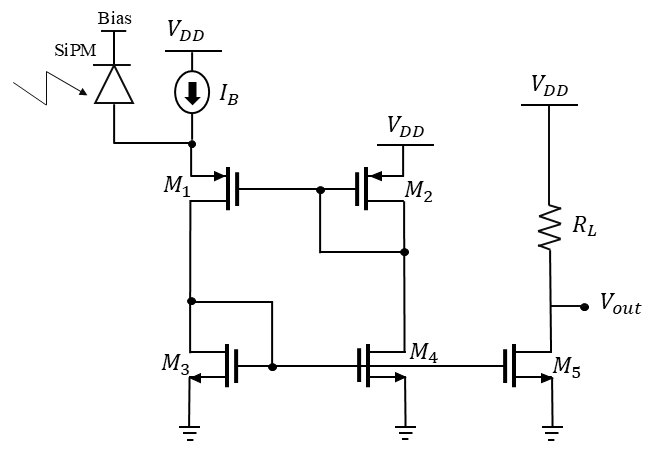}
\caption{Schematic of an amplifier with input impedance given by 
$R_{in} = \frac{1}{g_{m,1}} \left( 1 - \frac{g_{m,4}}{g_{m,2}} \cdot \frac{g_{m,1}}{g_{m,3}} \right)$.
It is evident that variations in the transistors' transconductance can result in a negative resistance, leading to instability. As a result, tuning is essential in this type of implementation. Additionally, the use of three branches to achieve lower impedance and convert current to voltage may lead to increased power consumption.}
\label{fig:CC}
\end{figure}

For the design of the preamplifier, the primary challenge was achieving very low impedance. The literature presents several methods for lowering impedance. For example, in \cite{sengupta2022sipm, santos2023negative, buonanno2021gamma, hosseini2020low, trigilio2018sipm, taghavi201510, belostotski2012wideband, chen2013cross, seifouri2015design}, the impedance can reach extremely low values, even approaching zero (Figure~\ref{fig:CC}). However, due to PVT (Process, Voltage, Temperature) variations, there is a risk of negative input impedance and instability. To mitigate this, a tuning section is required, which introduces additional power consumption, nonlinearity, and space constraints.

\begin{figure}[htbp]
\centering
\includegraphics[scale=0.4]{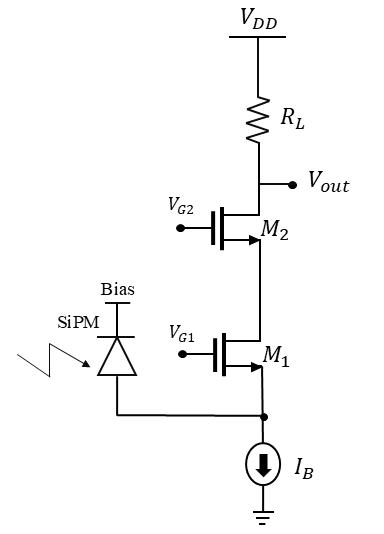}
\caption{Single-ended schematic of a low input impedance amplifier from NINO ASIC~\cite{calo2019sipm, gundacker2019high}, where the input impedance is given by $R_{in} \approx \frac{1}{g_{m1}}$. As shown, lowering the impedance requires increasing $g_{m1}$, which results in higher power consumption. Additionally, this design introduces an output swing limitation, which can result in saturation.}
\label{fig:powersimpl}
\end{figure}

Litruture also include a specific catagory, as demonstrated in \cite{shen2018silicon,dey2017cmos, anghinolfi2004nino,powolny2011time, chen2018diet}, achieving low impedance typically requires either increasing the transistor's transconductance or using a greater number of transistors. Both approaches, however, lead to higher power consumption (Figure~\ref{fig:powersimpl}).

\begin{figure}[htbp]
    \centering
    \includegraphics[scale=0.4]{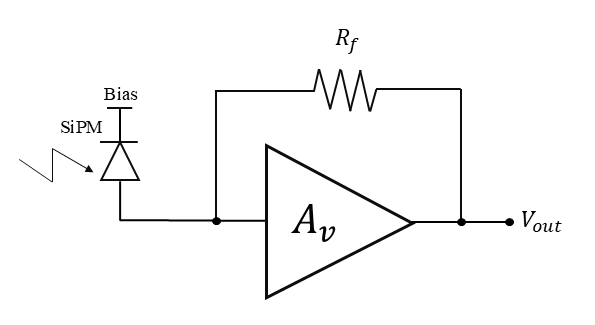}
    \caption{Schematic of a front-end with shunt feedback for reducing the input impedance. Here, $R_{\text{in}} = \frac{R_f}{1 + A_v}$, where $A_v$ is the amplifier gain. Additionally, the total gain is $R_f$. As evident, by decreasing $R_f$, the input impedance is reduced, which contradicts the increase in the total gain.}
    \label{fig:shunt}
\end{figure}

Another category of low-input impedance amplifiers is discussed in \cite{kim2010bandwidth, razavi2019transimpedance, ozkaya201764}, where there is often a trade-off between achieving low impedance and maintaining amplifier gain (Figure~\ref{fig:shunt}.

\begin{figure}[htbp]
\centerline{\includegraphics[scale=0.2]{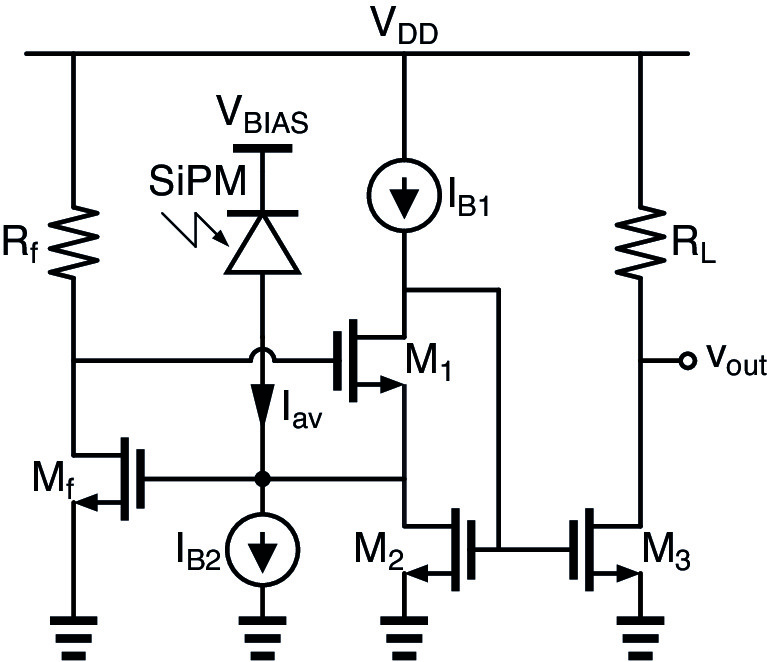}}
\caption{Schematic of the designed low input impedance preamplifier. $I_{av}$ is the avalanche current of the SiPM.}
\label{fig:amp}
\end{figure}

To avoid the mentioned obstacles in designing a low input impedance amplifier, such as instability, power consumption, and the dependency on gain, the amplifier shown in Figure~\ref{fig:amp} is designed. This design is primarily based on two key concepts for reducing input impedance using internal negative feedback.

The first feedback path is created by a flipped voltage follower (FVF)~\cite{carvajal2005flipped, ramirez2013flipped,akbari2024implementation, ramirez2023class, katourani2022ultra}, consisting of transistors $M_1$, $M_2$, $M_3$, and the current source $I_{B1}$.

The second feedback path, composed of $M_f$ and $R_f$, introduces a transconductance ($g_m$) boosting capability for $M_1$~\cite{yun20232pa, costanzo2018current,calo2019cmos, de2013regulated, rolo2012TOFPET, baghtash2011very, li2005g, orita2023time}. This configuration, sometimes referred to as a regulated cascode in the literature, further reduces the input impedance to:

\begin{equation}
    R_{in}\approx \frac{1}{g_{m1}g_{m2}R_{B1}(1+g_{mf}R_f)}
    \label{eq:RIN}
\end{equation}
being $R_{B1}$ the equivalent small-signal resistance of $I_{B1}$. 

As is noticeable, a negative term does not exist in Equation~\ref{eq:RIN}, and the input impedance can be further reduced by increasing $R_f$ and $R_b$, without requiring additional power consumption. Additionally, the DC gain can be defined as $v_{out}/i_{in} = N \cdot R_L$ (N is the size ratio between $M_3$ and $M_2$), which is independent of the input impedance.

To increase $N$ without increasing power consumption in the output branch or sacrificing output swing, a second bias current source, $I_{B2}$, with $I_{B2} < I_{B1}$, has been added. This approach allows the DC current of $M_2$ to remain low, while the size of $M_3$ can be increased without raising power consumption. At the same time, $M_1$ maintains a high transconductance, which is useful for reducing the input impedance, as the current $I_{B2}$ does not affect its drain current.

However, as described in~\cite{carvajal2005flipped}, the drain current of $M_2$ (i.e., $I_{B1} - I_{B2}$) cannot be set too low, as its reduced transconductance would negatively impact the input impedance.

Another observed challenge was the limitation in implementing a low-voltage amplifier and the reduction in output swing caused by transistor \(M_f\) (The output swing problem was indirectly due to the current mirror \(M_2\)-\(M_3\)) in the regulated cascode. Various methods to achieve low-voltage operation have been explored in the literature~\cite{michal2022regulated, ramirez2023class}. However, these approaches often require additional circuitry, resulting in increased power consumption, added complexity, and more complicated frequency behavior.

To overcome this challenge, \(M_f\) is selected from a low-threshold technology, which helps keep the transistors in the input branch in saturation under a low voltage supply. Additionally, choosing \(M_f\) from the low-threshold voltage library ensures that the drain-source voltage (\(V_{\text{DS}}\)) of \(M_2\), and consequently the \(V_{\text{DS}}\) of \(M_3\), remains low.

To achieve high-frequency performance, the size of the transistors in the pre-amplifier has been minimized. However, this introduces the challenge of DC mismatch between channels, preventing the use of a uniform threshold across all channels. To address this, a bypass capacitor is deliberately inserted between the pre-amplifier output and the comparator input. This capacitor provides DC decoupling between the two stages. As a result of this decoupling, the comparator's input can be biased using a diode-connected transistor, $M_b$ (Figure~\ref{fig:CH}). To reduce mismatch effects at this node, $M_b$ is made relatively large.

Besides the points mentioned in the last paragraph, there is another important reason for implementing a capacitor between the pre-amplifier and the comparator. According to~\cite{gola2013analog}, this capacitor can reduce the effect of dark counts on baseline shifts, which can degrade timing resolution. 

\begin{figure}[htbp]
    \centering
    \includegraphics[width=1.1\linewidth]{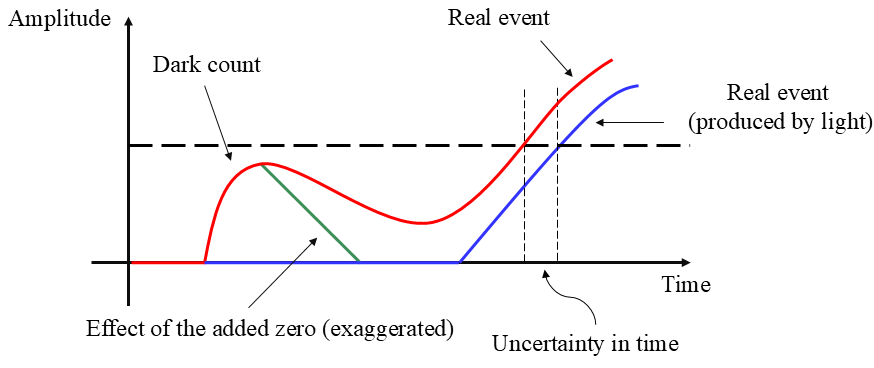}
    \caption{Illustration of the impact of dark counts in SiPMs on baseline shifts, resulting in jitter in timing measurements. Dark counts can cause baseline displacement, increasing uncertainty in the measurement of real events (referring to events generated by the SiPM in response to photon detection). To mitigate this effect, the capacitor acts as a high-pass filter, leading to a fast green decay that helps stabilize the baseline.}
    \label{fig:darkjiter}
\end{figure}

Dark counts in SiPMs are events where the detector generates a signal even without a photon being detected. These events are caused by thermally generated carriers and their occurrence is completely random. Due to the SiPM's long decay time, dark counts can cause random shifts in the baseline. When a real event occurs, this baseline shift introduces uncertainty and jitter in the timing measurement (Figure~\ref{fig:darkjiter}).

By using a bypass capacitor, which introduces a zero to counteract the pole responsible for the decay time, the baseline is stabilized, thereby improving timing accuracy.




Concerning the frequency compensation of the preamplifier, it was initially not a primary concern. Since there is no designed feedback loop, the possibility of phase shift leading to instability was not considered problematic. Worthy of notice, in both the PCB and ASIC, the layout of the ground and power rails was designed to be as wide as possible to avoid any unwanted feedback. However, as mentioned earlier, two internal loops exist, which could potentially cause instability. The risk of instability from these loops was investigated using the methodology highlighted in reference~\cite{ramirez2004low}.

Nevertheless, given the large size of the SiPM input capacitor, the pole generated by this capacitor is the dominant pole, and it occurs at a lower frequency compared to the other poles. It is sufficiently distanced from them, ensuring that the phase of the frequency response is not adversely affected, making the implementation of frequency compensation circuits unnecessary. Additionally, implementing frequency compensation arrangements could compromise the preamplifier’s fast performance characteristics and limit its bandwidth. The primary goal, as discussed in Section~\ref{sec:XRY}, is to maintain a circuit with a fast frequency response suitable for ToF-CT applications.

\begin{figure}[htbp]
\centerline{\includegraphics[scale=0.14]{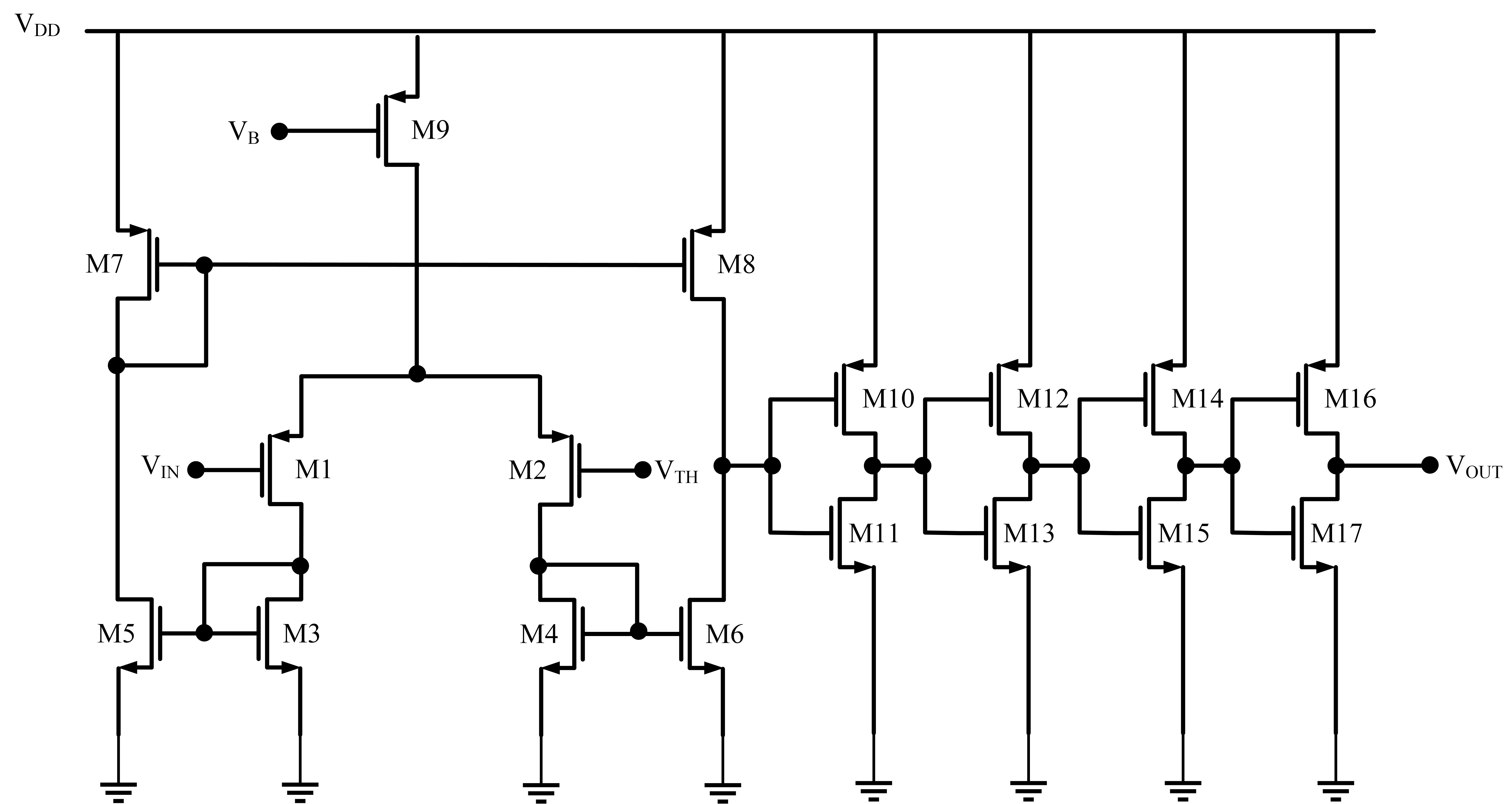}}
\caption{Transistor-level schematic of the voltage comparator, which consists of a single-ended op-amp and a series of inverter chains at the output. In addition to enhancing gain, the comparator generates a digital signal based on a threshold. The width and rising time of this digital signal play an important role in determining the timing performance of the design. This digital signal is then fed into the later stages of data acquisition.
}
\label{fig:comp}
\end{figure}

Figure~\ref{fig:comp} presents the schematic of the voltage comparator, which consists of a single-ended operational amplifier and an inverter chain, primarily designed to significantly increase the gain. The comparator’s threshold is adjustable and derived externally, allowing for flexible threshold settings. For the next generation of the ASIC, a programmable and tunable threshold is planned to be introduced. This enhancement will improve the chip's functionality, especially when integrated into a detector chain with a large number of channels.

\begin{figure}[htbp]
\centerline{\includegraphics[scale=0.35]{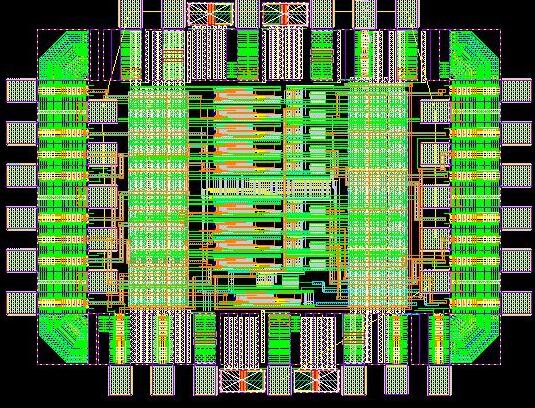}}
\caption{Layout of LUCAS 
including pad rings - each channel has an area of $263\,\textup{µm} \times 40\,\textup{µm}$.}
\label{fig:LAY}
\end{figure}

Figure~\ref{fig:LAY} illustrates the layout of the tape-out chip, which was fabricated using TSMC's 65nm low-power CMOS technology, operating with a 1.2V power supply (VDD). The chip includes eight primary channels designed specifically for SiPM readout, ensuring efficient signal processing. The total area occupied by the chip is 1mm by 1mm, reflecting a compact design that integrates all the necessary functionalities within a minimal footprint.


\begin{figure}[htbp]
\centerline{\includegraphics[scale=0.15]{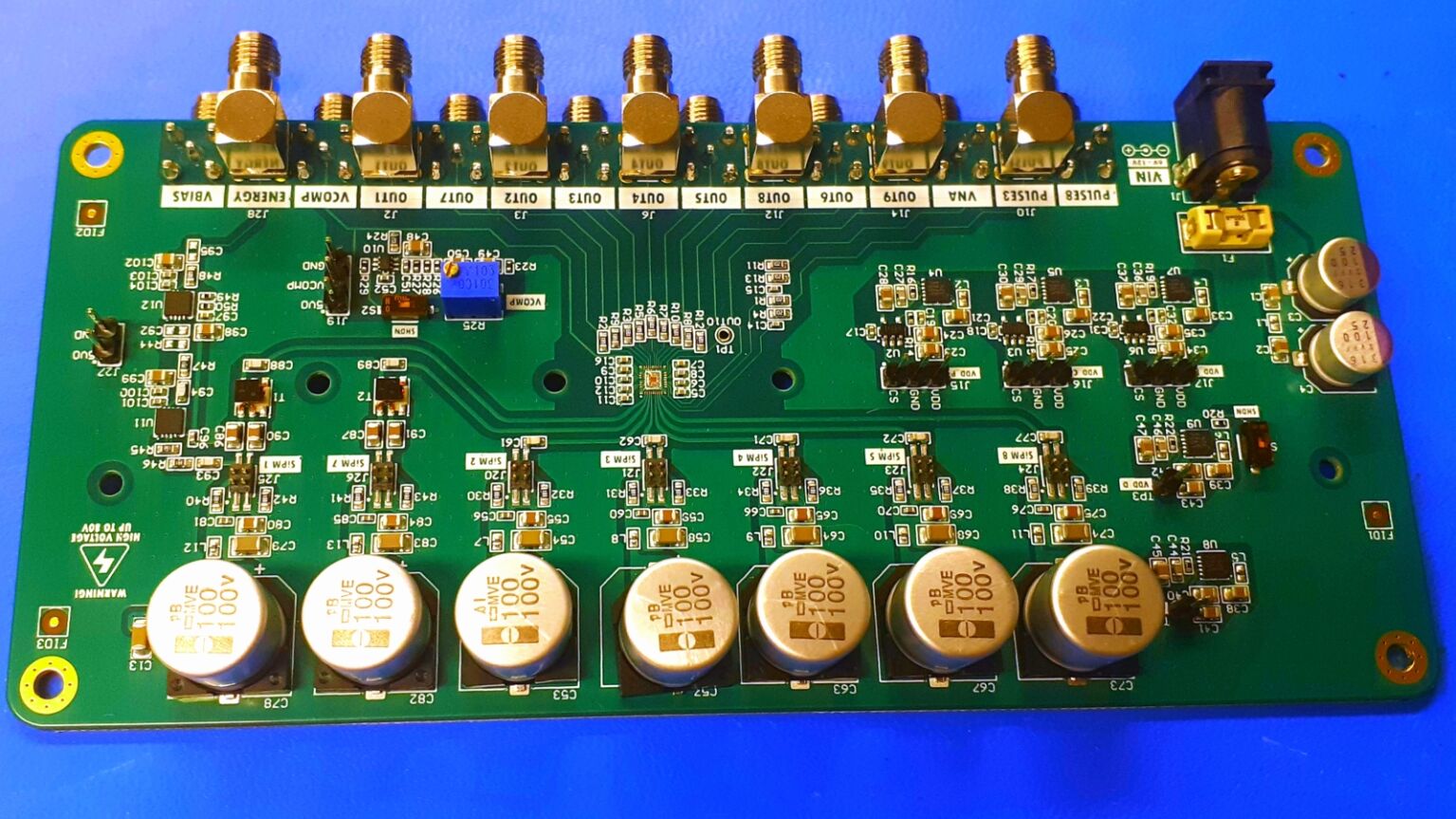}}
\caption{The LUCAS test board is designed to offer various functionalities, such as enabling a high voltage bias for the SiPM and allowing the threshold for the ASIC to be adjusted.}
\label{fig:Board}
\end{figure}

Figure~\ref{fig:Board} shows the wire-bonded chip mounted on the test board, which has been specifically designed to support and enable various desired functionalities. The test board provides the necessary platform for performing comprehensive tests and measurements, ensuring that all required operations of the chip can be evaluated efficiently.

The subsequent sections provide a detailed overview of the materials and methods employed to characterize the tape-out chip. These sections explain the various measurements and techniques used to assess the chip's performance, ensuring it meets the required specifications for the ToF-CT project. This includes outlining key steps taken to evaluate the chip’s characteristics, as well as the approaches implemented to ensure accuracy and reliability in the measurements.

\section{Materials \& Methods}
\label{sec:MM}

This section introduces the materials and methods used to measure, validate, and verify the ASIC specifications. First, the approach for measuring input impedance is discussed. Next, the procedures employed for measuring and confirming the single photon time resolution and X-ray time of flight, along with the resources utilized in these processes, are presented.

\subsection{The Input Impedance}
\label{sec:IMPEDANCE}

In this study, the input impedance plays a pivotal role, particularly due to its primary goal to negate the parasitic effects arising from the interconnections between SiPM cells, a concept elaborated upon in Section \ref{sec:ASIC}. To facilitate input impedance measurements, 
a Keysight Technologies P9374A vector network analyzer (VNA) was employed to measure the input impedance of the ASIC. This equipment was chosen for its 26.5 GHz operational frequency range, which facilitates measurements at high frequencies.

For an accurate assessment of the input impedance of the preamplifier, a dedicated channel has been integrated, tailored specifically for evaluating the input impedance. This strategy allows the Keysight P9374A VNA to be directly connected to the channel's input (bypassing the circuitry designed for SiPM bias and the interconnections between the SiPM and ASIC on the board). 

To further enhance measurement accuracy and eliminate the influence of the PCB traces on impedance measurements, three calibration PCBs (bare PCBs without any components) have been prepared with terminations of 0 ohms, 50 ohms, and an open circuit. These calibration boards are separately utilized in conjunction with Keysight's software by connecting each board to the VNA and setting the appropriate calibration profile. Following this, the LUCAS PCB is connected,
and data for the $S_{11}$ parameter is acquired within the 5 GHz frequency bandwidth. A Python script was then used to convert the $S_{11}$ parameter to impedance and to plot figures showing the real, imaginary, and amplitude components of the impedance at each measured frequency. 

The outcomes of the impedance measurements are comprehensively presented in Section~\ref{sec:RES}. This section elaborates on the findings, highlighting the key characteristics of the preamplifier's input impedance across various frequencies, providing a foundation for further analysis and comparison with relevant literature.

\subsection{Single Photon Time Resolution}
\label{sec:SPTR}

\begin{figure}[htbp]
    \centering
    \includegraphics[width=0.5\textwidth]{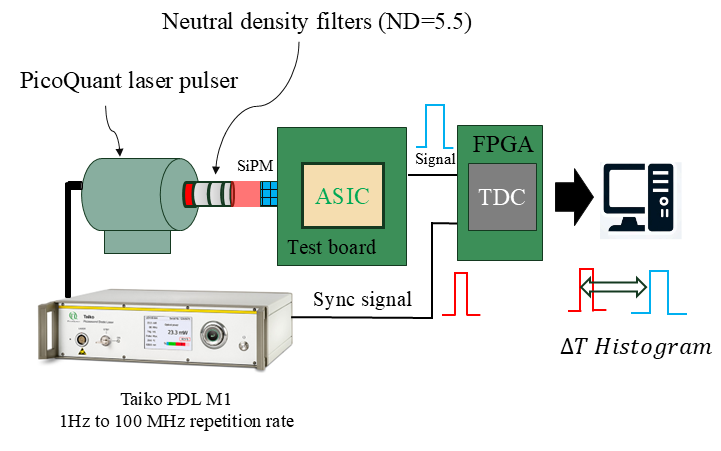}
    \caption{Configuration of the single photon time resolution detection setup, including the pulsed laser source, neutral density filters, SiPM, the ASIC, and the acquisition system. The interconnection board is omitted for simplicity.}
    \label{fig:SPTR}
\end{figure}

The configuration for Single Photon Time Resolution (SPTR) detection, shown in Figure \ref{fig:SPTR}, incorporates a PicoQuant laser diode for the pulsed laser source, a series of neutral density filters, an AFBR-S4N22P014M 2$\times$2 mm$^2$ Broadcom SiPM \cite{Broadcom2024SiPMArray}, the LUCAS chip, an interconnection board, and a Xilinx ZC706 FPGA which acts as a data collector, all aiming to accurately detect single photons. 

The PicoQuant LDH-I series, with Taiko PDL M1 module, emits pulses with 60 ps $<$ FWHM at the wavelength of 420 nm, which is in the wavelength range that the employed SiPM has the highest photon detection efficiency. 


To ensure a single-photon acquisition setup, a series of neutral-density filters were used to reduce the laser beam's intensity to the single-photon level.

 The employed SiPM has a breakdown voltage of 32.5V and a maximum overvoltage of 16V. It also has a terminal capacitance of 160 pF.




 An interconnection board was utilized to interface properly between the ASIC and FPGA, providing differential signals through an LVDS link to feed the FPGA \cite{bouchard2017low}.

\begin{figure}[htbp]
    \centering
    \includegraphics[width=0.43\textwidth]{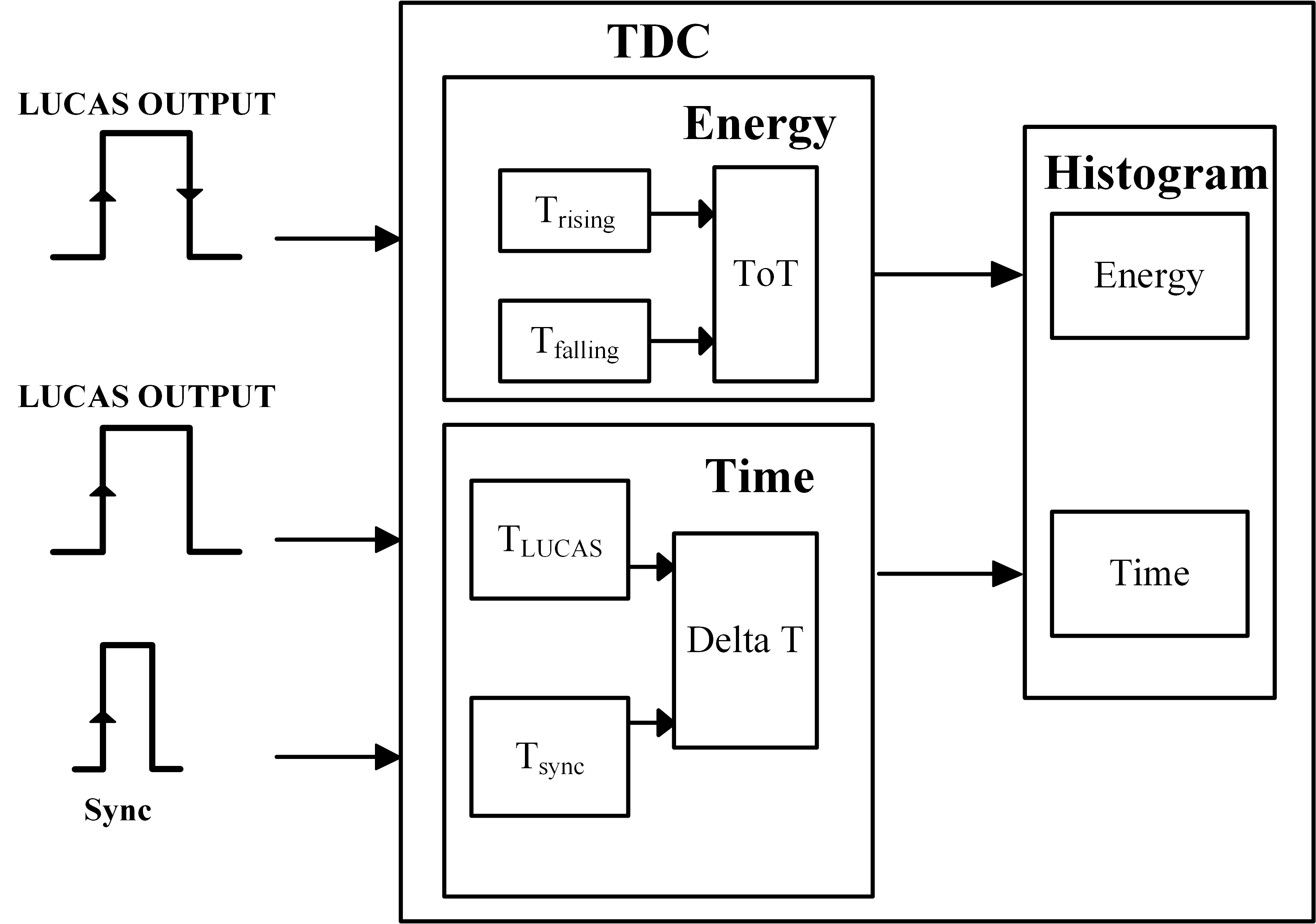}
    \caption{High-level diagram of the TDC. The TDC utilizes a delay line technique for measuring $\Delta T$ with a 20 ns time window and 10 ps resolution. It can also perform ToT measurements for energy quantification and photon counting, with 1 ns binning and 8 manually defined energy bins.}
    \label{fig:TDC}
\end{figure}

To collect the data, a multiple-tapped delay lines Time-to-Digital Converter (TDC) embedded in a Xilinx ZC706 FPGA was used \cite{wingering2024fpga}. A high-level diagram of the TDC is shown in Figure~\ref{fig:TDC}.

For time resolution measurement, the TDC employs a delay line technique to measure $\Delta T$, where $\Delta T$ is the time difference between the start and stop signals—specifically, the sync signal (used as a reference) and the output of the ASIC. The TDC is configured to measure $\Delta T$ within a fixed time window of 20 ns, with a time resolution of 10 ps.

As shown in Figure~\ref{fig:TDC}, the TDC not only measures time resolution but also captures Time over Threshold (ToT) for energy measurement or photon counting. This is achieved using an auxiliary board mounted on the FPGA board, which splits the output signal for both time and energy measurements. The TDC then creates histograms of the data with 1 ns binning, corresponding to its energy resolution. Additionally, the TDC can histogram $\Delta T$ data into 8 manually defined energy bins, enabling the capture of events such as single-photon or two-photon occurrences.


To observe the spectrum of detection for one, two, and three-photon events, the ASIC threshold and SiPM biases were calibrated using dark counts.  This method allowed setting the threshold at values sufficiently above the baseline noise, yet close enough to avoid exceeding the amplitude of a single photon.

\begin{figure}[htbp]
    \centering
    \includegraphics[width=0.5\textwidth]{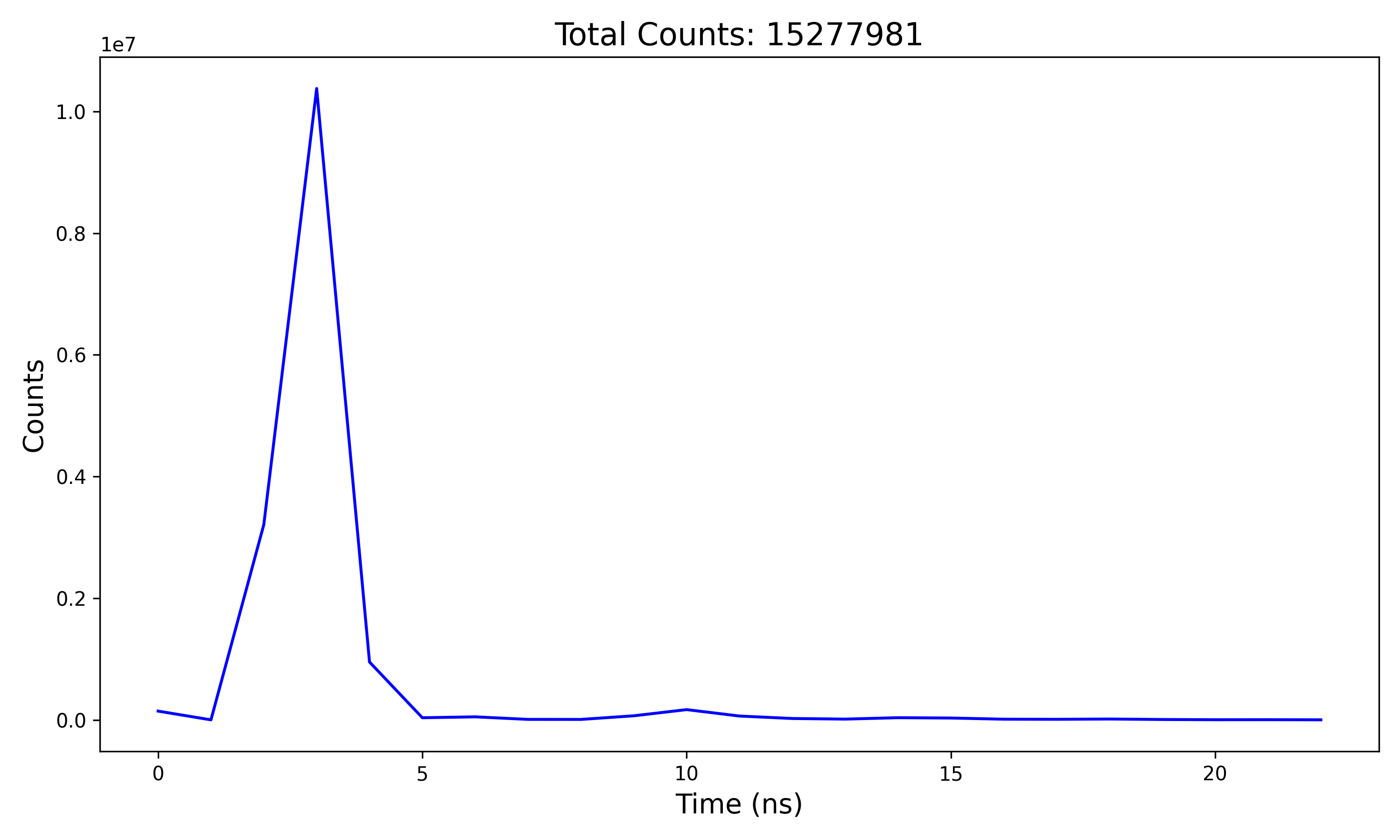}
    \caption{In a 30-second acquisition, with the bias of 40 for the SiPM, the total number of dark counts for the 2 mm by 2 mm SiPM with a dark count rate of \(125 \, \text{kcounts/sec/mm}^2\) is given by: 
    \(125 \, \text{kcounts/sec/mm}^2 \times 4 \, \text{mm}^2 \times 30 \, \text{sec} = 15,000,000 \, \text{counts}.\)}
    \label{fig:dark}
\end{figure}

To determine the Time-over-Threshold for a single photon, measurements from dark counts were utilized. This means the data was collected without any laser trigger. The ToT data was extracted, and a histogram was plotted (Figure~\ref{fig:dark}). The first peak in the histogram was assumed to represent the ToT for a single SPAD firing, which corresponds to a single photon.

To verify that this peak indeed represents a single SPAD firing, two methods were applied. First, the number of dark counts over a 30-second acquisition time (\(t_{\text{acq}}\)) was calculated using information from the SiPM datasheet. According to the datasheet, with a bias voltage of 40 V, the SiPM with a size of 2 mm by 2 mm ( \(A_{\text{SiPM}}\)) and a dark count rate (\(R_{\text{dark}}\)) of \(125 \, \text{kcounts/sec/mm}^2\) should exhibit a total dark count of:

\begin{align}
N_{\text{dark}} &= R_{\text{dark}} \times A_{\text{SiPM}} \times t_{\text{acq}} \nonumber \\
&= 125 \, \text{kcounts/sec/mm}^2 \times 4 \, \text{mm}^2 \times 30 \, \text{sec} \nonumber \\
&= 15,000,000 \, \text{counts}
\label{eq:dark_count}
\end{align}

The sync signal frequency was adjusted to a high resolution of 100 MHz during the measurement to sample as many dark events as possible. As shown in Figure~\ref{fig:dark}, the measured number of dark counts closely aligns with the calculated value, confirming that the first peak in the histogram is not an artifact or noise.

The second method of verification is based on the low likelihood of more than two SPADs firing simultaneously in dark mode. Given this low probability, the large first peak in the histogram is reasonably assumed to represent the firing of a single SPAD. This peak can thus be used to reliably estimate the ToT for a single photon.

Building on this analysis, the maximum threshold for detecting a single photon was identified by gradually increasing the threshold until the system could no longer detect a single SPAD firing. This was determined by observing when the first peak, corresponding to single SPAD events, disappeared and was replaced by multiple SPAD firing events. The disappearance of the peak was noted through changes such as the shift of the peak to a lower ToT value, followed by a sudden drop in the count. The process was then repeated to determine the minimum overvoltage required to detect a single photon.


After identifying the time-over-threshold for single-photon events within the spectrum using the methodology described above, the TDC was programmed via FPGA to extract the time histogram data for the selected duration. Following this, the laser intensity was set to the minimum value of 760 nW, with a pulsing frequency of 2 MHz. This frequency ensures that the pulsing rate exceeds the dark count rate.



\subsection{X-ray Time of Flight}
\label{sec:XRY}

\begin{figure}[htbp]
    \centering
    \includegraphics[width=1\linewidth]{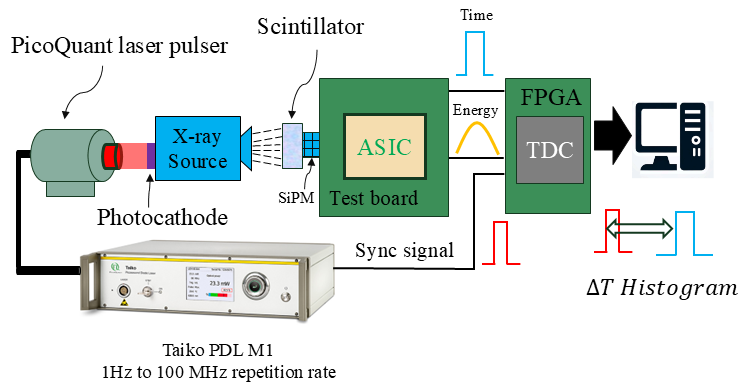}
    \caption{The setup for X-ray Time-of-Flight (ToF) measurement includes a pulsed X-ray source with its laser trigger, a scintillator, an SiPM, an ASIC, and an acquisition system (FPGA). For simplicity, the interconnection board is excluded.}
    \label{fig:xrayset}
\end{figure}

The setup for X-ray Time-of-Flight (ToF) measurement, shown in Figure~\ref{fig:xrayset}, includes a pulsed X-ray source, a scintillator for converting X-ray photons to visible light, an AFBR-S4N22P014M 2$\times$2 mm$^2$ Broadcom SiPM that acts as a light detector, the LUCAS chip as a front-end for signal processing, an interconnection board for interfacing, and a Xilinx ZC706 FPGA, which serves as the data collector.

The X-ray source used in this setup is manufactured by Hamamatsu and operates at a maximum voltage of 120 kVp. The generation of X-rays is triggered by the Piquant pulsed laser, which activates the source. Each time a laser pulse strikes the photocathode, it excites the material, causing the emission of electrons. These emitted electrons are then accelerated towards a tungsten target under high voltage.

Upon reaching the tungsten target, the electrons undergo rapid deceleration as they interact with the electric field of the tungsten nuclei. This sudden deceleration causes the electrons to lose energy, which is released in the form of X-ray photons through a process known as Bremsstrahlung radiation (German for 'braking radiation'). During this process, the change in the electron's trajectory due to the interaction with the electric field results in the emission of electromagnetic radiation across a continuous spectrum.

As a result, in this setup, the X-rays are emitted as pulses, each with a pulse width of less than 100 ps (FWHM), providing high temporal resolution for the measurement. The X-ray source can accommodate average tube currents of up to 10 µA. The tube current refers to the average electrical current flowing through the X-ray tube during operation, determining the number of electrons emitted by the photocathode and subsequently accelerated towards the target.

A scintillator is required in the setup to convert X-ray photons into visible light. For this purpose, two different scintillators were selected to be tested separately to determine which one provides better performance. A \(4 \,\text{mm} \times 4\,\text{mm} \times 3\, \text{mm}\) cerium-calcium co-doped LYSO scintillator from Taiwan Applied Crystals and a \(4\,\text{mm} \times 4\,\text{mm} \times 0.43\,\text{mm}\) 600 nm InGaN/GaN MQW scintillator made by Crytur were employed, both paired with a \(2\,\text{mm} \times 2\,\text{mm}\) Broadcom SiPM.

\begin{figure}[htbp]
    \centering
    \includegraphics[width=1\linewidth]{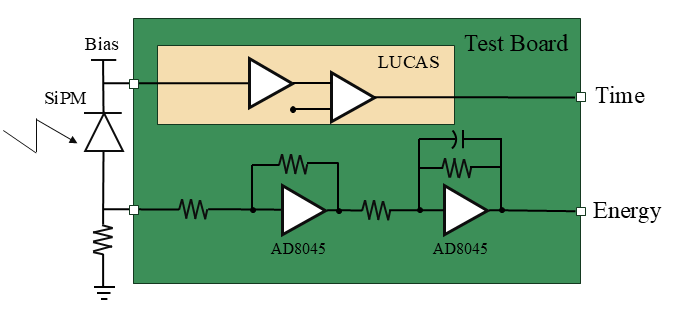}
    \caption{The AC model of the discrete energy channel, used for measuring X-ray energy and obtaining the X-ray spectrum, includes an amplifier and an integrator.}
    \label{fig:EnergyCH}
\end{figure}

Energy measurement is essential in time-of-flight X-ray systems, primarily to distinguish system noise from X-ray signals and to assess the time resolution of X-rays across different parts of the spectrum. This information is crucial for improving image reconstruction results and for further investigation of ToF CT as a new module. Due to the fast output of the preamplifier, there is a risk of saturation in the time-over-threshold (ToT) measurement at high energies. Therefore, a discrete energy channel has been used to integrate the energy (Figure~\ref{fig:EnergyCH}), and the ToT method is then applied to the output of this channel for accurate energy measurement.

An interconnection board, similar to the one described in Section~\ref{sec:SPTR}, was used as an interface between the ASIC and FPGA. Through this interface, the energy and time signals were fed to the TDC, allowing for the extraction of two types of histograms: the delta T histogram (measuring the time difference between sync and time signals) and the energy histogram (based on Time over Threshold, using the energy signal information).

\begin{figure}[htbp]
    \centering
    \includegraphics[width=1\linewidth]{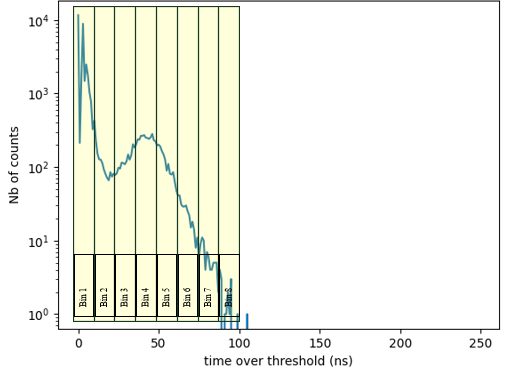}
    \caption{The TDC has the capability to define multiple bins based on Time over Threshold, with adjustable bin sizes. Delta T histograms can then be extracted from specific regions of the spectrum corresponding to these defined bins. The data presented is for a Crytur scintillator operating with a 44V bias on the SiPM.}
    \label{fig:8binxray}
\end{figure}

The TDC configuration, along with its histogramming capability, as described in Section~\ref{sec:SPTR}, allows for the definition of 8 bins based on Time over Threshold (Figure~\ref{fig:8binxray}). Using these defined ToT bins, the TDC can extract delta T histograms corresponding to each bin. This functionality aids in assessing the timing performance across different parts of the spectrum, enabling more detailed investigations into X-ray behavior and Time-of-Flight Computed Tomography. Additionally, it helps filter out noise and unwanted data, focusing on the timing performance of X-ray-related signals.

To accurately convert Time-over-Threshold measurements to energy, a calibration procedure was essential due to the inherent non-linear relationship between ToT and energy. This non-linearity arises because the ToT method does not linearly scale with the amount of energy detected, which can introduce distortions in the measured energy spectrum. Therefore, a fitting process was used to correct these distortions and ensure more precise energy measurements.

Four monoenergetic radioactive sources (Am-241, Co-57, Ge-68, and Cs-137) were used to obtain reference data. The ToT values for these sources were measured across different SiPM bias voltages and comparator thresholds to capture how the response varies under different operating conditions. Changing the bias voltage affects the gain and sensitivity of the SiPM, which in turn alters the ToT measurements for the same energy. By measuring ToT across a range of bias voltages and thresholds, a more comprehensive set of calibration data was created, mapping ToT to energy. This information allowed for a more accurate calibration that can account for non-linearities in the ToT-energy relationship, ensuring reliable energy measurements across different settings.

To establish the relationship between ToT and energy, a model function was selected based on the expected behavior of the detector system. The chosen expression was designed to accurately account for the non-linearities observed in the data:
\begin{equation}
\text{ToT}(E; a, b, c, d) = a + b \cdot \ln(E + d) + c \cdot \ln^2(E + d)
\label{eq:tot_energy}
\end{equation}
where:
\begin{itemize}
    \item \( a \) is a constant offset,
    \item \( b \) scales the logarithmic term,
    \item \( c \) accounts for additional non-linearities,
    \item \( d \) prevents issues at low energies by shifting the input range.
\end{itemize}

The parameters \( a \), \( b \), \( c \), and \( d \) were determined by fitting the function to the collected data using a \textit{least-squares optimization} approach~\cite{chen2021gamma}. The objective of the optimization was to minimize the total sum of squared residuals \( S \), defined as:
\begin{equation}
S(a, b, c, d) = \sum_{i=1}^{N} \left[\text{ToT}_{\text{measured}, i} - \text{ToT}_{\text{model}, i}(E_i; a, b, c, d)\right]^2
\label{eq:sum_of_squares}
\end{equation}
where:
\begin{itemize}
    \item \( \text{ToT}_{\text{measured}, i} \) represents the experimentally measured ToT for the \( i \)-th energy value,
    \item \( \text{ToT}_{\text{model}, i} \) is given by Equation \ref{eq:tot_energy}, using the parameters \( a \), \( b \), \( c \), and \( d \),
    \item \( N \) is the total number of data points.
\end{itemize}

The minimization process involves calculating the \textit{partial derivatives} of \( S \) with respect to each parameter:
\begin{equation}
\frac{\partial S}{\partial a}, \quad \frac{\partial S}{\partial b}, \quad \frac{\partial S}{\partial c}, \quad \frac{\partial S}{\partial d}
\end{equation}
These derivatives indicate how sensitive the sum of squared errors \( S \) is to changes in each parameter. By using these derivatives, the fitting algorithm can determine the direction and magnitude of adjustments needed for each parameter, iteratively reducing \( S \) to find the optimal values of \( a \), \( b \), \( c \), and \( d \).

A \textit{Jacobian matrix} \( J \) is constructed to contain these partial derivatives:
\begin{equation}
J = \begin{bmatrix}
\frac{\partial \text{ToT}_{\text{model}, 1}}{\partial a} & \frac{\partial \text{ToT}_{\text{model}, 1}}{\partial b} & \frac{\partial \text{ToT}_{\text{model}, 1}}{\partial c} & \frac{\partial \text{ToT}_{\text{model}, 1}}{\partial d} \\
\frac{\partial \text{ToT}_{\text{model}, 2}}{\partial a} & \frac{\partial \text{ToT}_{\text{model}, 2}}{\partial b} & \frac{\partial \text{ToT}_{\text{model}, 2}}{\partial c} & \frac{\partial \text{ToT}_{\text{model}, 2}}{\partial d} \\
\vdots & \vdots & \vdots & \vdots \\
\frac{\partial \text{ToT}_{\text{model}, N}}{\partial a} & \frac{\partial \text{ToT}_{\text{model}, N}}{\partial b} & \frac{\partial \text{ToT}_{\text{model}, N}}{\partial c} & \frac{\partial \text{ToT}_{\text{model}, N}}{\partial d}
\end{bmatrix}
\label{eq:jacobian}
\end{equation}

The optimization algorithm iteratively adjusts the parameters, using the information from the partial derivatives to minimize \( S \). The process stops when the changes in \( S \) are small enough to indicate that the best-fit parameters have been found, i.e., when \( S \) reaches a local minimum.

After fitting, the calibrated function was used to convert ToT values into energy, enabling the accurate reconstruction of the X-ray energy spectrum.

\section{Results}
\label{sec:RES}

\subsection{Input Impedance}

\begin{figure}[htbp]
\centerline{\includegraphics[scale=0.3]{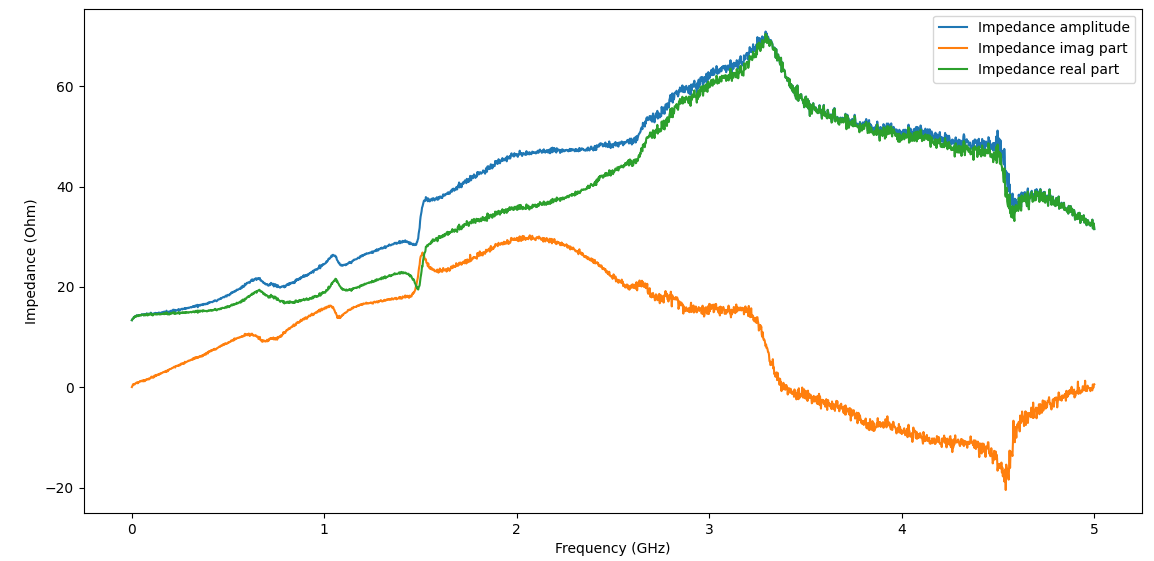}}
\caption{Impedance characteristics of LUCAS preamplifier across frequency range}
\label{fig:s11}
\end{figure}

Figure \ref{fig:s11} shows the measured acquisition of impedance across a frequency range up to 5 GHz. Notably, for frequencies below 3.5 GHz, the input impedance of the preamplifier remains below 50 ohms. It is important to highlight that some research \cite{kratochwil2021roadmap, pourashraf2022electronic}, that uses discrete components for measuring time resolution, often employs RF amplifiers with a 50-ohm input impedance. These amplifiers typically utilize technologies such as SiGe (silicon-germanium) bipolar transistors or other similar high-performance semiconductor technologies. In contrast, this design is implemented in CMOS technology, which limits both unity current gain frequency ($f_{T}$) and transistors' intrinsic transconductance. The effect of the latter is pronounced in Equation~\ref{eq:RIN}.

\subsection{Single Photon Time Resolution}

\begin{figure}[htbp]
    \centering
    \begin{subfigure}{0.45\textwidth}
        \centering
        \includegraphics[width=\textwidth]{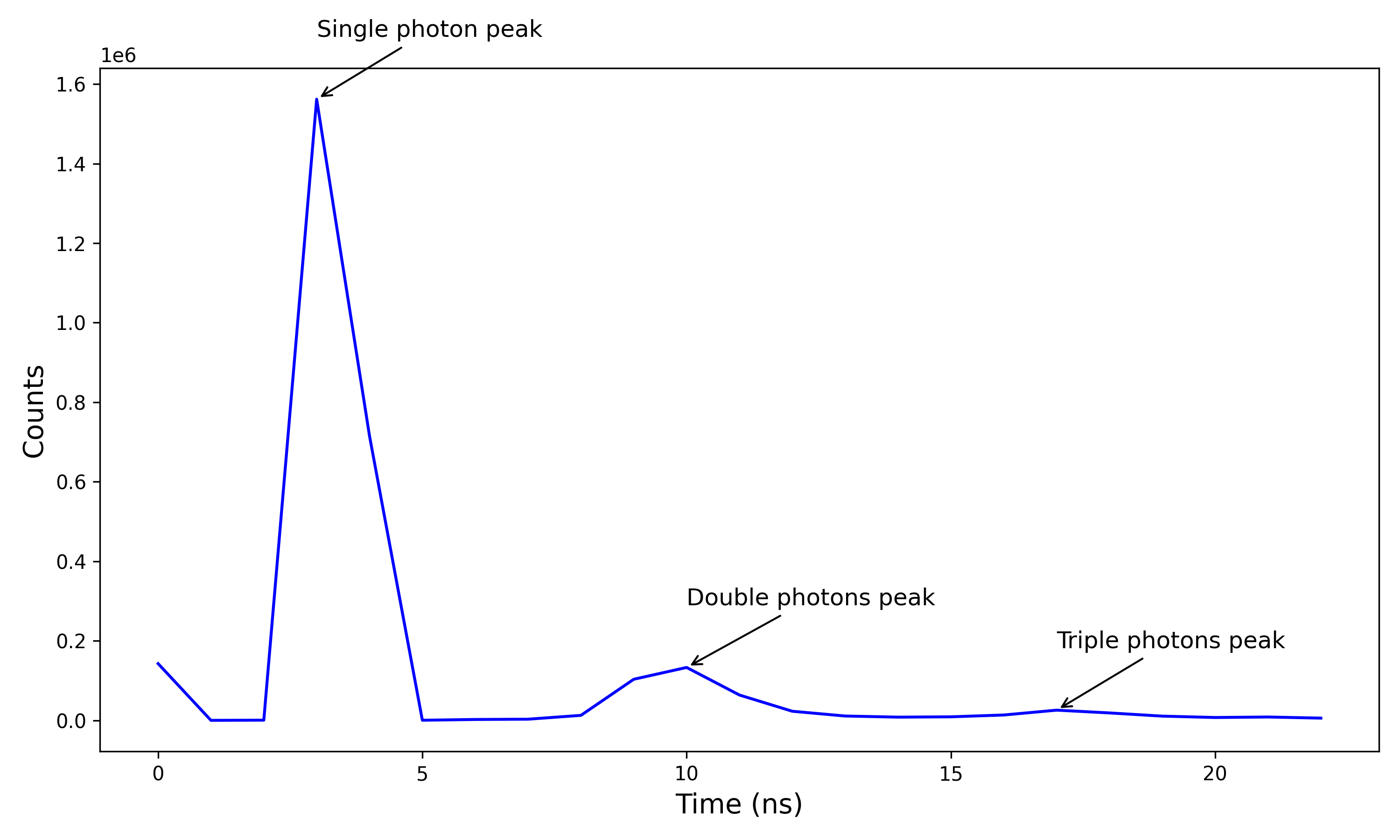}
        \caption{Photon counting histogram where photons are distinguished and discriminated based on the Time over Threshold (ToT). The first peak represents single-photon incidents.}
    \end{subfigure}
    \hfill
    \begin{subfigure}{0.45\textwidth}
        \centering
        \includegraphics[width=\textwidth]{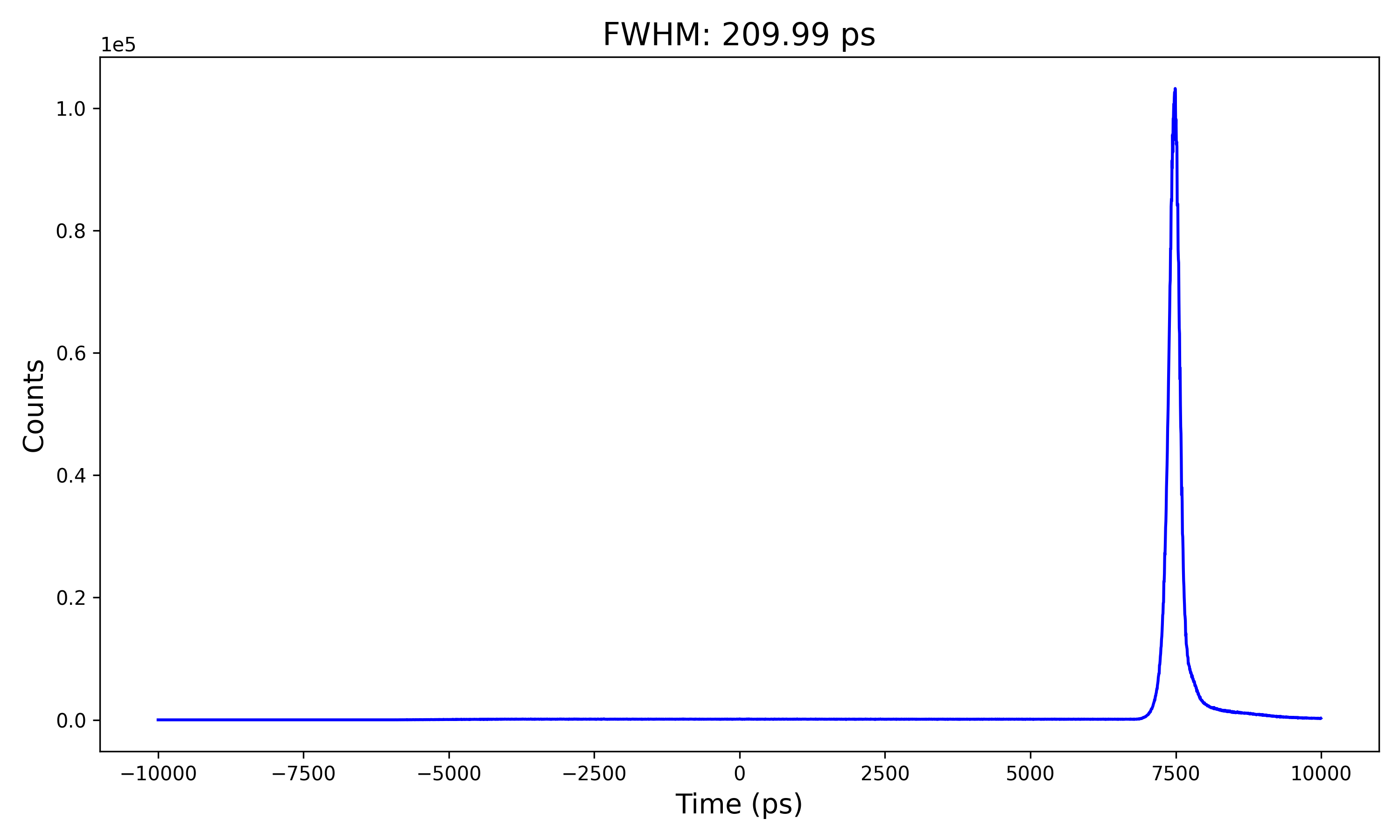}
        \caption{Delta T histogram of the photons in the first peak. The histogram is based on the time difference between the sync signal (used as a reference) and the ASIC output.}
    \end{subfigure}
    \caption{Photon counting spectrum and Single Photon Time Resolution results obtained under the following conditions: a SiPM bias voltage of 40V, an ASIC threshold of 646 mV, a laser power intensity of 760 nW, and a pulsing rate of 2 MHz.}
    \label{fig:sptrres}
\end{figure}

Figure~\ref{fig:sptrres} shows the photon counting spectrum and the delta T histogram from the SPTR measurement. The methodology, acquisition process, and verification approaches are detailed in Section~\ref{sec:SPTR}. This result was obtained with a SiPM bias of 40V and a threshold of 646 mV on the ASIC, demonstrating an SPTR of 210 ps FWHM.

\begin{figure}[htbp]
\centerline{\includegraphics[scale=0.4]{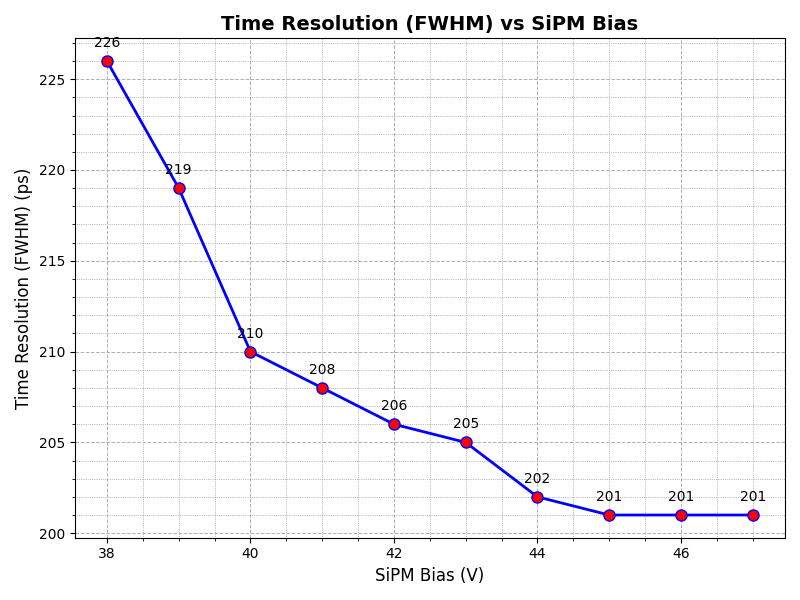}}
\caption{Summary of SPTR measurements for varying SiPM biases at the 646 mV threshold on the ASIC. The results demonstrate that increasing the SiPM bias enhances SPTR performance, as higher bias results in greater gain, faster signals, and reduced susceptibility to jitter.}
\label{fig:SPTRvsBIAS}
\end{figure}

Furthermore, SPTR measurements were conducted for different SiPM biases. Figure~\ref{fig:SPTRvsBIAS} summarizes the results, showing that increasing the SiPM bias improves the SPTR. This was expected, as higher bias increases the SiPM gain, leading to faster signals with higher amplitude, which are less susceptible to jitter.

Single photon time resolution can serve as a key figure of merit for comparing different SiPM readout systems. Table~\ref{tab:comparison} presents a comparison of the results from this study with those reported in other works. Additionally, as certain parameters such as the SiPM size and power consumption are not reflected in the SPTR alone, these parameters have also been included in Table~\ref{tab:comparison} to provide a more comprehensive understanding of the factors influencing performance improvements.

\begin{table}[htbp]
    \centering
    \begin{tabular}{|l|c|c|c|}
        \hline
         \textbf{ASICs} & \textbf{Power} & \textbf{SPTR (FWHM)} & \textbf{SiPM Size} \\
        & \textbf{[mW/ch]} & \textbf{[ps]} & \textbf{[mm\(^2\)]} \\
        \hline
        LUCAS & 3.2 & 201 & 2 x 2 \\

        HRFle x ToT\cite{sanchez2021hrflextot} & 3.5 & 167 & 3 x 3 \\
        
        FastIC\cite{gomez2022fastic} & 12 & 140 & 3 x 3 \\
        NINO\cite{nemallapudi2016single} & 27 & 160 & 3 x 3 \\
        TOFPET2\cite{nadig2022comprehensive, bugalho2019experimental} & 8 & 195 & 3 x 3 \\
        
        STiCi\cite{munwes2015single} & 25 & 158 & 1.3 x 1.3 \\
        Liroc\cite{saleem2023study} & 3.3 & 169 & 2 x 2 \\
        Petiroc 2A\cite{ahmad2018petiroc2a} & 6 & 200 & 3 x 3 \\
        \hline
    \end{tabular}
    \caption{Comparison of SPTR (FWHM), power consumption, and SiPM size for various ASICs, including LUCAS.}
    \label{tab:comparison}
\end{table}








\section{Conclusion}
\label{sec:CON} 

This paper introduces LUCAS, a novel low-power, ultra-low jitter ASIC designed for Silicon Photomultiplier (SiPM) readout in Time-of-Flight Computed Tomography (ToF-CT) applications. ToF-CT represents a transformative approach to X-ray imaging by exploiting precise timing information to distinguish scattered photons from primary photons, thereby reducing noise and artifacts in medical imaging. The success of this technique hinges on the timing performance of the underlying electronics, particularly the readout system. LUCAS addresses this critical need by combining an innovative low-input impedance preamplifier and an optimized comparator within a compact and efficient architecture. 

One of the key challenges in SiPM readout is mitigating the effect of parasitic capacitance, which introduces a low-frequency pole, limiting the timing resolution. LUCAS employs a current-mode preamplifier with dual internal feedback paths to achieve exceptionally low input impedance without sacrificing power efficiency or stability. Unlike traditional designs that require higher power consumption or face instability risks, LUCAS ensures robust performance through a combination of novel circuit techniques, including flipped voltage followers and regulated cascodes, implemented in TSMC’s 65nm CMOS technology. The result is a preamplifier with an input impedance below 50 ohms up to 3.5 GHz, offering significant advantages over RF amplifiers commonly used in discrete setups.

The ASIC demonstrated excellent timing resolution during experimental validation, achieving a Single Photon Time Resolution (SPTR) of 201 ps FWHM at 3.2 mW/channel, which compares favorably with state-of-the-art ASICs. The SPTR performance improves with higher SiPM bias voltages, as demonstrated by systematic measurements, further highlighting the robustness of the LUCAS design under varying operating conditions. 

In addition to timing performance, LUCAS supports energy measurements through Time-over-Threshold (ToT) techniques. A calibration process using monoenergetic radioactive sources established a precise mapping of ToT to energy, accounting for non-linearities in the detector system. This capability is particularly crucial for ToF-CT, where energy discrimination aids in filtering noise and enhancing image quality. The integration of an energy channel with an additional amplifier and integrator ensures accurate energy capture, even at high input rates, avoiding saturation while maintaining high resolution.

The compact design of LUCAS, occupying just 1 mm² of chip area, demonstrates its potential for scalability in multi-channel systems. The 8-channel architecture, combined with a low power footprint, makes LUCAS a viable solution for large-scale ToF-CT systems requiring high-density integration without compromising energy efficiency. The inclusion of bypass capacitors and advanced biasing techniques further ensures stable operation, even under challenging conditions, such as dark count noise and baseline shifts commonly encountered in SiPM readouts.

From an application perspective, LUCAS addresses critical impediments in ToF-CT imaging, including noise reduction, energy discrimination, and the ability to scale to larger systems. Its low power consumption ensures feasibility for portable or high-density imaging systems, while its ultra-low jitter enables precise photon timing required for advanced imaging modalities. The ASIC also highlights the potential of CMOS technology in achieving performance metrics that rival or exceed those of specialized RF or discrete solutions.

Future work will focus on enhancing the LUCAS design to include features such as programmable thresholds, improved energy resolution, and expanded channel counts. These enhancements aim to further optimize its applicability in ToF-CT and extend its use to other photon-counting applications, such as positron emission tomography (PET) and high-energy physics experiments. Additionally, exploring advanced calibration algorithms and machine learning techniques for real-time noise filtering and energy mapping could significantly improve the overall performance of ToF-CT systems.

In conclusion, LUCAS represents a significant advancement in SiPM readout electronics, achieving a balance of high timing resolution, low power consumption, and scalability. Its innovative circuit design, validated through comprehensive experimental testing, underscores its potential to enable next-generation ToF-CT systems with improved image quality, reduced noise, and lower radiation doses. By addressing fundamental challenges in SiPM readout, LUCAS paves the way for broader adoption of ToF-CT and other photon-timing technologies in clinical and research settings.


%

\appendices



\ifCLASSOPTIONcaptionsoff
  \newpage
\fi

\end{document}